\documentclass[12pt,preprint]{aastex}
\usepackage{natbib}

  \shortauthors{Jourdain et al.}
\shorttitle{Polarized  emission in Cygnus X-1  with \textit{INTEGRAL} SPI}

\begin{document}

 \title{SEPARATION OF TWO  CONTRIBUTIONS TO THE HIGH ENERGY EMISSION OF CYGNUS X-1:
POLARIZATION MEASUREMENTS WITH \textit{INTEGRAL} SPI 
\footnote{Based on observations with INTEGRAL, an ESA project with instruments and science 
data centre
funded by ESA member states (especially the PI countries: Denmark, France, Germany, Italy, 
Spain, and Switzerland), Czech Republic and Poland with participation of Russia and USA.}}

\author{E. Jourdain\altaffilmark{1}, J. P. Roques\altaffilmark{1}, M. Chauvin\altaffilmark{1}  
and D. J. Clark\altaffilmark{1,2}  }
\altaffiltext{1}{Universit\'e de Toulouse; UPS-OMP; IRAP;  Toulouse, France\\ CNRS; IRAP; 9 
Av. colonel Roche, BP 44346, F-31028 Toulouse cedex 4, France} 
\altaffiltext{2}{Createc Ltd; Unit 8, Derwent Mill Commercial Park, Cockermouth, Cumbria, CA13 
0HT, United Kingdom}

\author{\it Received  ; accepted  }


\begin{abstract}
 Operational since 2002 on-board the INTEGRAL observatory, the SPI spectrometer 
 can be used to perform polarization measurements 
in the hard X-ray/soft$\gamma$-ray domain ($\sim$130 keV - 8 MeV). 
 However, this phenomenon is complex to measure at high energy and requires high fluxes.
 Cyg X-1 appears
 as the best candidate amongst the X-ray binaries since it is one of the brightest persistent
sources in this energy domain. Furthermore, a polarized component has recently been reported
above 400 keV from IBIS data. We have therefore dedicated our efforts to develop the required 
tools
to study the polarization in the  \textit{INTEGRAL} SPI data and have first applied them to
2.6 Ms of Cyg X-1 observations, covering 6.5 years of the INTEGRAL mission.
 
We have found that the high energy emission of Cyg X-1 is indeed polarized, with 
a mean polarization fraction  of 76$\%$ $\pm$ 15$\%$  at a position angle estimated to 
42$^\circ$
$\pm$ 3$^\circ$, for energies above 230 keV. The polarization fraction clearly increases with 
energy. In the 130-230 keV band, the polarization fraction is lower than 20 $\%$, but exceeds 
75 $\%$ between 370 and 850 keV, with  the (total) emission vanishing above 
this energy.  This result strongly suggests that the emission originates from
the jet structure known to emit in the radio domain. The same synchrotron process could be 
responsible
for  the emission  from radio to MeV, implying the presence of high energy electrons.   This 
illustrates why the polarization of the high energy emission
in compact objects is an increasingly important 
observational objective.
\end{abstract}      


\maketitle

\section{INTRODUCTION}
The high energy emission of X-ray binaries remains  a  debated topic.
Between a few and 100-200 keV, a thermal inverse Comptonization of disk photons in a 
hot corona, accompanied by a disk reflection component, provides a good description of the 
data, but
we are far from a complete understanding.
In particular, some bright sources exhibit an excess of emission above 200-300 keV.
This feature has been confirmed by several instruments operating above 300 keV, 
first in Cyg X-1 (see for example Nolan et al. 1981 or 
Liang \& Nolan, 1984) but also in a few other bright hard X-ray sources 
(i.e. GX 339-4 or the X-ray transient GRO J0422+32, see Bouchet et al., 1993; 
Johnson et al., 1993; Roques et al., 1994). The COMPTEL/CGRO instrument has shown 
that, in Cyg X-1, this component extends up to a few MeV, with a power law slope around 2.6 or 
3, depending on the source state (McConnell et al., 2002).
 This is often interpreted as a Comptonization process but with a non-thermal electron 
population, possibly linked to the thermal population.
Alternative models  based on jet emission also exist (see for instance Markoff et al. 2001). 
Indeed, it is clear now that many X-ray binary systems harbor (relativistic)
variable radio jets and that radio and X-ray emissions are linked (Gallo et al. 2003).

Polarization measurements are a powerful tool to get specific information 
on the emission mechanisms and related parameters, as an effective supplement to spectral and 
timing studies.
While common in radio or optical bands, 
 they are less accessible in the high energy domain.  
However, two instruments aboard the INTEGRAL observatory  allow such measurements.
The IBIS imager obtains the polarization information from the Compton scattering
occuring between its two separate detector planes (Forot et al. 2007). 
In comparison, the SPI spectrometer obtains the polarization from the scattering of photons 
within its large volume 
detector plane composed of 19 independent germanium crystals. 

First measurements have been performed by both instruments independently 
on the Crab pulsar (Dean et al, 2008; Forot et al, 
2008)  
and a strong gamma-ray burst, GRB041219A (McGlynn et al. 2007; G\"otz et al. 2009), 
demonstrating that
polarization information can and must be sought in the high energy domain. 

Cyg X-1 is, with the Crab, one of the  main sources in the hard X-ray/soft $\gamma$-rays sky. 
It has been observed in many occasions by various instruments in this domain and presents
  complex spectral variability around two main spectral states (Jourdain et al. 2012 and 
references therein). It is therefore an
obvious candidate for high energy polarization measurements. Laurent et al. (2011) recently
published the detection of polarized emission, leading to the most popular scenarios
regarding the high energy emission of black hole X-ray binaries to be significantly revisited.  
These polarization studies are expected to provide extra information to decipher the 
mechanisms of the central engine.

In this paper, we use  SPI observations to investigate the  polarized
emission in the high energy spectrum of Cyg X-1.  
Section 1 describes  the instrument, analysis principles  and data set.
We then present the results obtained and conclude by discussing the first constrains
we can learn from them.
 
\section{INSTRUMENT, DATA ANALYSIS AND OBSERVATIONS}
The SPI  spectrometer aboard the  INTEGRAL observatory consists of a coded mask 
associated with a detector plane of 19 germanium crystals, imaging the sky
between 20 keV and 8 MeV. The large field of view (30$^\circ$ in diameter), coupled with a 
limited number of pixels (19),
results in the use of a dithering strategy for observations to increase the number of 
independant measurements taken of a region of sky. An observation becomes a series of 
exposures (or science windows, SCW) of 30-40 minutes duration, taken in a predetermined 
pattern of pointings, which changes how the mask shadow is projected on to the detector plane. 
The pointing direction between SCWs usually varies by $\sim 2.3^\circ$. 
Two or three observations (dedicated to different targets) are successively 
performed during one INTEGRAL revolution  (or orbit). Each revolution lasts $\sim$ 3 days,
resulting in $\sim2.5$ days of uninterrupted operating mode, when excluding the crossing 
of the radiation belts.
A description of the SPI telescope as well as its global  performance can be found in Vedrenne 
et al.
 (2003) and Roques et al. (2003).
 
To use the polarimetry capability of SPI requires going deeper into the detection process and
 corresponding data analysis than is usually required. 
When a photon interacts with the camera, it loses its energy via Compton scattering and/or
 photoelectric absorption.
When the incident photon deposits energy within more than one pixel 
this is classified as a "multiple event", as opposed to a "single event" where the energy is 
deposited within a single pixel. It is the angular distribution of the scatter directions 
involved in the multiple events that allows the determination of the polarization information.
Indeed, the Klein-Nishina cross section for a linearly polarised photon
presents itself as a function of the polarisation angle (more precisely of cos$^2\eta$ ),
 with its maximum at  $\eta$ = 90$^{\circ}$. 
However, due to the competition between photoelectric and Compton processes, the multiple events
play a significant role only for photons  above $\sim$ 130 keV, where they represent 
$\sim  20\%$  of the detected photons for the SPI geometry.

Events are tagged as "multiples" when they occur in (at least) 2 detectors, within a 
time-window of 350 ns.
For these events, we are able to reconstruct the initial energy by adding the measured 
deposits, while
the scattering direction is given by the identification of the involved detectors. The   
imaging information is contained here in 42 pairs of adjacent detectors (called 
pseudo-detectors), which play the same role as the 19 individual crystal for the single 
events. 
 For any observation, the events are extracted in a similar way than for the single events 
(Jourdain et al. 2009), but
using the imaging  transfer matrix for the multiple events.  The background shape is 
determined  
during empty field observations, and its normalization factor is allowed to vary 
during fitting on a six hours timescale.
The source flux is considered constant on the revolution (ie $\sim$ 1-3 days) timescale. The 
source and background fluxes are obtained for each energy band through a model fitting 
procedure 
convolving the sky model with  the  transfer matrix.

By carrying out this procedure in a number of channels, we can build the spectrum of the 
source,
based on the multiple events. Similarly to the standard (single events) data procedure,  
 the count spectrum is related to the incident photon spectrum through the appropriate
(ground calibrated, see Atti\'e et al., 2003) redistribution matrix, used in the spectral 
deconvolution step. This pipeline has been first validated using Crab Nebula
observations. The total spectrum obtained for the Crab with $\sim$ 600 ks of data has been 
found to be perfectly compatible with the result
deduced from the standard analysis chain between 20 keV and 3 MeV, making us confident in our
analysis method and response matrices. 

The response matrices used in the spectral analysis have been built for an unpolarized
 emission. However, the instrument response varies with the polarization of the incident flux, 
since the polarization introduces an anisotropy in the scattered photons. We have therefore   
replaced the  unpolarized (ie averaged) matrix response with a polarized response in any study 
aiming to recover
 the polarization information, such as this.   
A dedicated paper (Chauvin et al., submitted) explains in detail the polarimetry technique
used in the present analysis, which we summarize here.

An incident polarized photon preferentially scatters perpendicular to its polarization vector. 
This scattering forms a probability distribution that requires the observation of many photons 
to determine the original polarization vector.  

 For the polarization analysis, the imaging transfer function is computed at each incident 
direction and for each polarization angle, thanks to a detailed SPI mass model based on the 
Geant4 software.  \\
 For a given pointing, simulations are carried out for 18 polarization angles ($0-170^\circ$ 
in 
$10^\circ$ steps, 100$\%$ polarized) 
and one unpolarized. To introduce a partial polarization fraction, a set of data is built by 
mixing the polarized and unpolarized simulated distributions: 
\begin{equation}
G4(f_p,\Pi)=\frac{f_p \times G4(\Pi)}{100}+\frac{(100-f_p) \times G4(\varnothing)}{100}
\label{Percent_equ}
\end{equation}
 where $G4(f_p,\Pi)$ represents the (Geant4) simulated data for a $f_p\%$ polarized flux at an 
angle $\Pi$ , $G4(\Pi)$ is
the $100\%$ polarized simulated data at angle $\Pi$ and $G4(\varnothing)$ the unpolarized 
simulated data. 

For each exposure, we compare the observed data and the simulated patterns, allowing two free 
parameters: the source flux and the background normalizations. They are determined through a 
linear least squares resolution which  minimizes the expression:

\begin{equation}
     \chi^2 (f_p,i) = \sum  (M_{f_p,i}(E)-D(E))^2/D(E)^2 
    \label{chi2_equ}
 \end{equation}
\begin{equation}
M_{f_p,i} =x \times G4({f_p},\Pi)+y \times B
\label{Source_equ}
\end{equation}
where x is the source normalization, G4,  the simulated count distribution, y is the 
background 
normalization and B,the background spatial distribution, over the energy band considered (E). 
The source model $G4_{sd}(f_p,\Pi)$ 
describes the number of counts  as a function of 
source polarization fraction, $f_p$, and angle, $\Pi$.\\ 
In a global data analysis, all the selected pointings are processed simultaneously, for each 
polarization
 angle and $f_p$ value, to give a final 18 $\times$ 101 $\chi^2$ array.
Among them, the lowest value identifies the most probable (polarization angle, polarization 
fraction) parameter values. \\
The errors on the 2 free parameters have also been estimated through Monte-Carlo simulations
with 1000 iterations for each configuration. The obtained distributions are 
 centered around the simulated values (angle or fraction) and their widths correspond to 
values given by the $\chi^2$
statistics criterion $\Delta\chi^2$ = $\chi^2$ + 1, which has been used for the remainder of 
the analysis.
More details on the method, Monte-carlo simulations, analysis procedure and validation tests
are given in a forthcoming paper (Chauvin et al., submitted). Note that our simulation code is  
based on an improved version of the code used in previous polarization studies (McGlynn et al. 2007,
 Dean et al. 2008), while the data analysis procedure has been developed independently and includes 
specific parameters, for the background handling for exemple.
 We have started our Cyg X-1 analysis with the same dataset presented for the spectral 
analysis of Jourdain et al. (2012), which covers 2003 June - 2009 December period. In spite of 
an increase of the flux by more than a factor two or three and a clear spectral variability, 
the source was mainly in 
its hard state, with some incursions into the intermediate ones. 
Since the multiple events concern  only $\sim$ 20$\%$ of the total flux, the polarization 
analysis suffers
from much weaker statistics. Moreover, for a given incident direction, the evolution of the 
projected pattern with the polarization
angle is small. This implies that the sky model must be as simple as possible (only one 
source),
and leads us to exclude some periods where the transient source EXO 2023+375, close to Cyg 
X-1,
flares. Revolutions presenting a too high normalization background have also been removed. 
Finally, we present here the analysis of a data  set limited to 2.6 Ms of effective time, 
corresponding to  28 INTEGRAL revolutions, distributed over 6.5 years (see Tab. 1).
 
\section{RESULTS}
 
\begin{figure}[h]
\begin{center}$
\begin{array}{cc}
\includegraphics[width=2.4in]{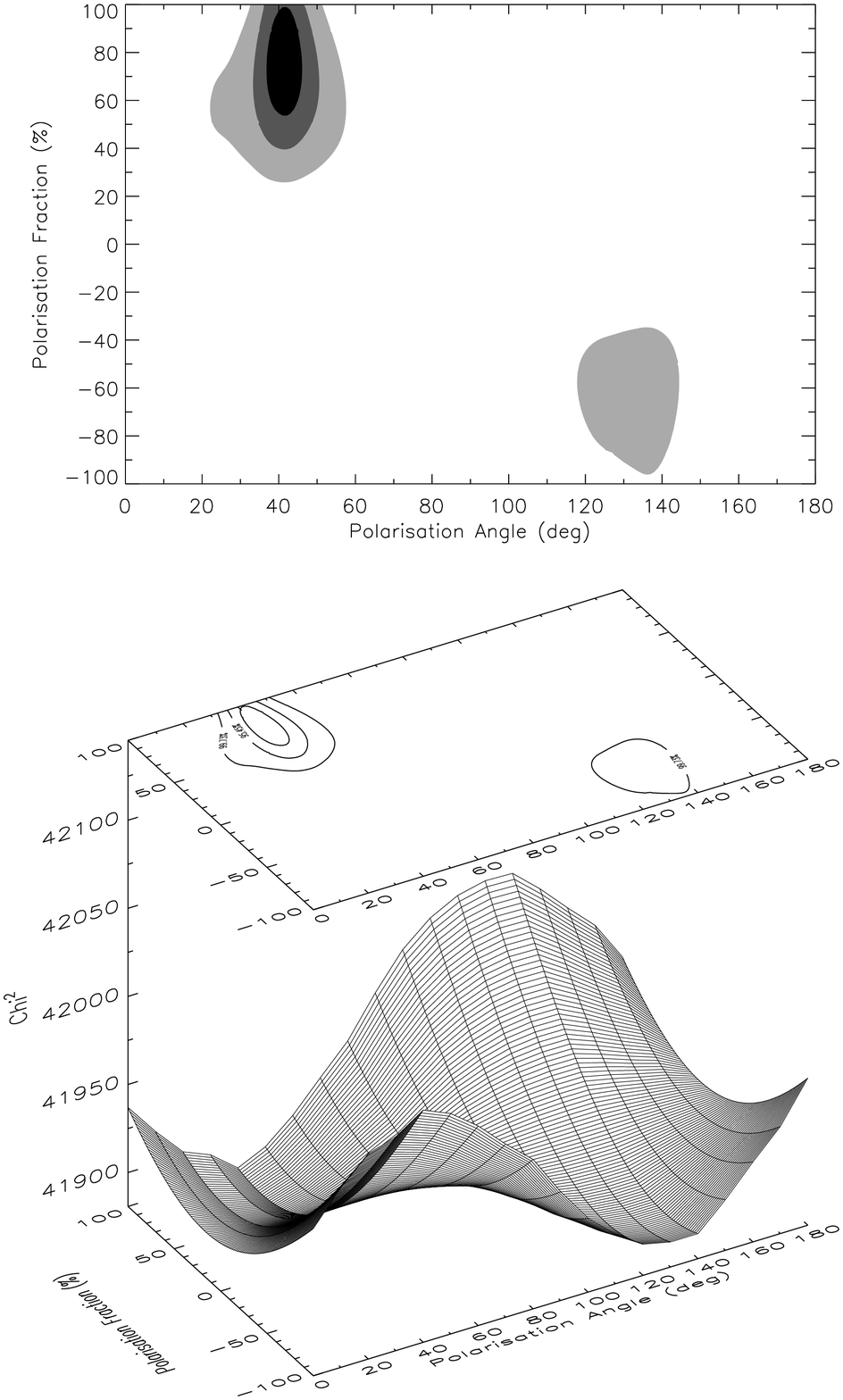}&
\put(-180,140){a)}
\put(-180,290){b)}
\put(-5,290){d)}
\put(-5,140){c)}
\includegraphics[width=2.5in]{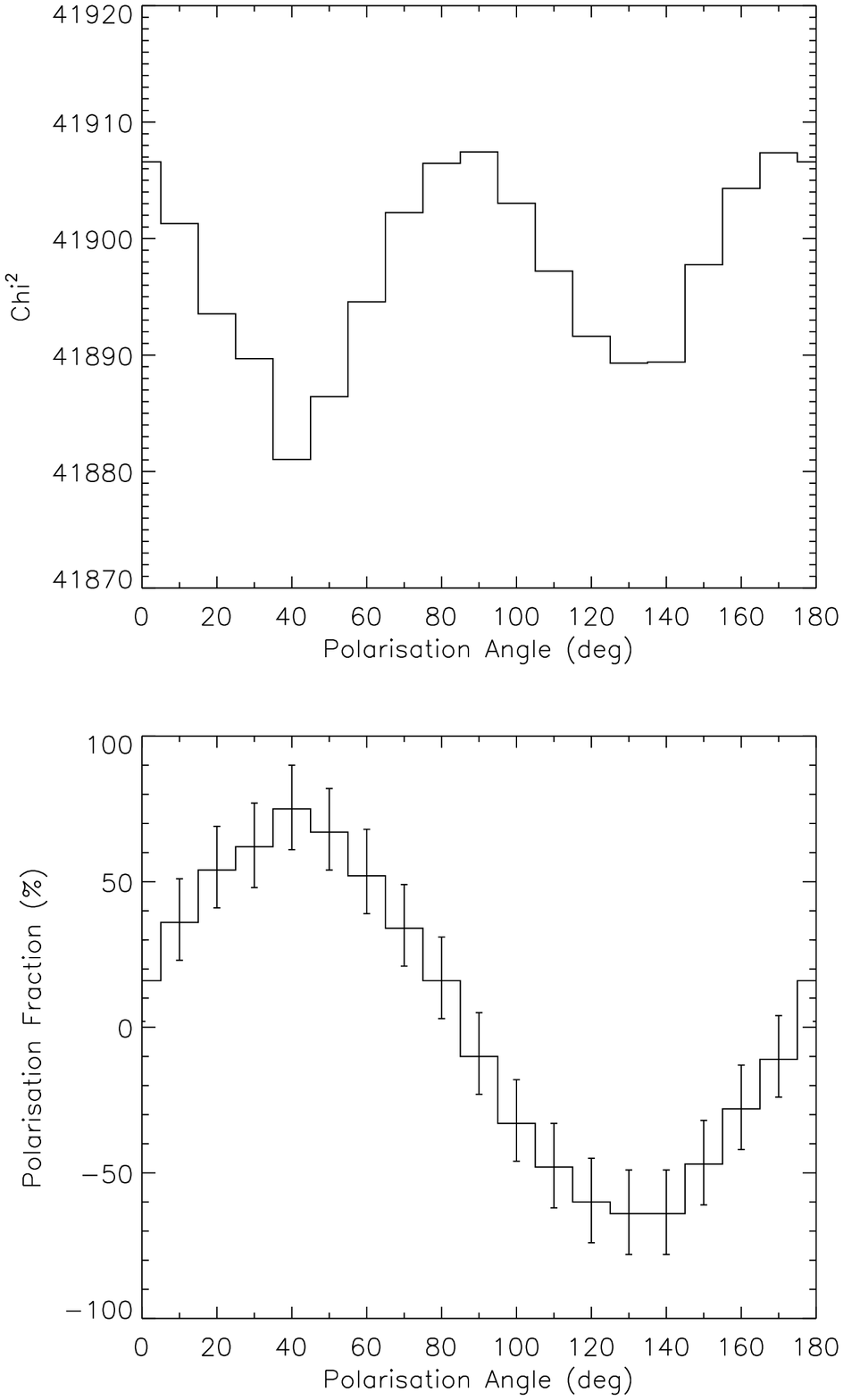}\\
\end{array}$
\end{center}
\caption{Left panel: a) $\chi^2$ map result versus position angle and polarization fraction 
in the 230-850 keV band. b) The contour projections at $\Delta\chi2$=2.3, 6.2, 11.8.
 Right  panel: c) Polarization fraction giving the best fit for each
 position angle and d) the corresponding $\chi^2$ curve.}
 \label{fig:mean}

\end{figure}

We analyzed the total dataset presented in Table 1, and we initially considered 
the energy band from 230 to 850 keV.
The photons above 850 keV have not been included since the flux in this energy range appears 
very weak in the spectral analysis of the source (Jourdain et al. 2012) and decreases the 
significance of the result. As described above, for a given energy band, we compare the 
spatial distribution of the photons 
recorded on the detector plane and in the simulation, letting the flux of the source and 
background
to vary on the revolution and  6 hours timescales, respectively.
The results are presented in Figure \ref{fig:mean}. The displayed surface corresponds to
the $\chi^2$ values obtained from eq. (1), in the (position  angle, polarization fraction) 
space.
Additionally, we also plot (right panel) the best  polarization fraction 
 (given by the lowest $\chi^2$ value) for each of the tested polarization angles. 
 The  $\chi^2$ surface appears smoothly structured with a trough, identifying
 the best fit between the observations and the simulated data. In the top projection, a 
contour plot 
 illustrates more clearly the $\chi^2$ behavior around its minimum.
 The three grey level areas are associated to the 1 (2 and 3) $\sigma$ levels,
 as deduced from the $\chi2$= $\chi2$ + 2.3 (+ 6.2, + 11.8 resp.) formula.

We see that the observations are better described by an incident emission with 
a polarization fraction of 76\% $\pm$ 15\% at a position angle of 42$^\circ$ $\pm$ 3$^\circ$.

\begin{figure}[h]
\begin{center}$
\begin{array}{cc}
\includegraphics[width=2.4in]{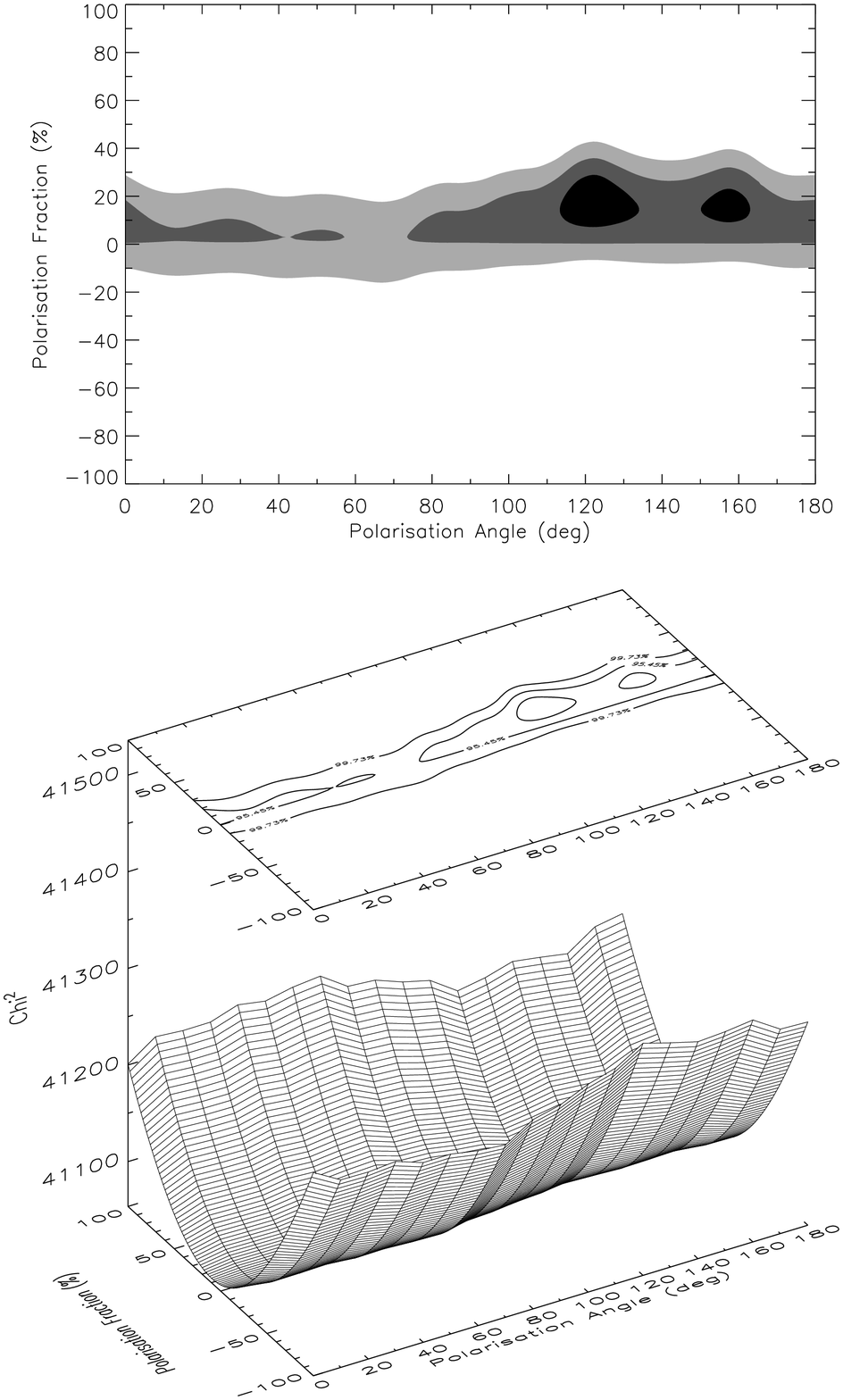}&
\put(-180,140){a)}
\put(-180,290){b)}
\put(-5,290){d)}
\put(-5,140){c)}
\includegraphics[width=2.5in]{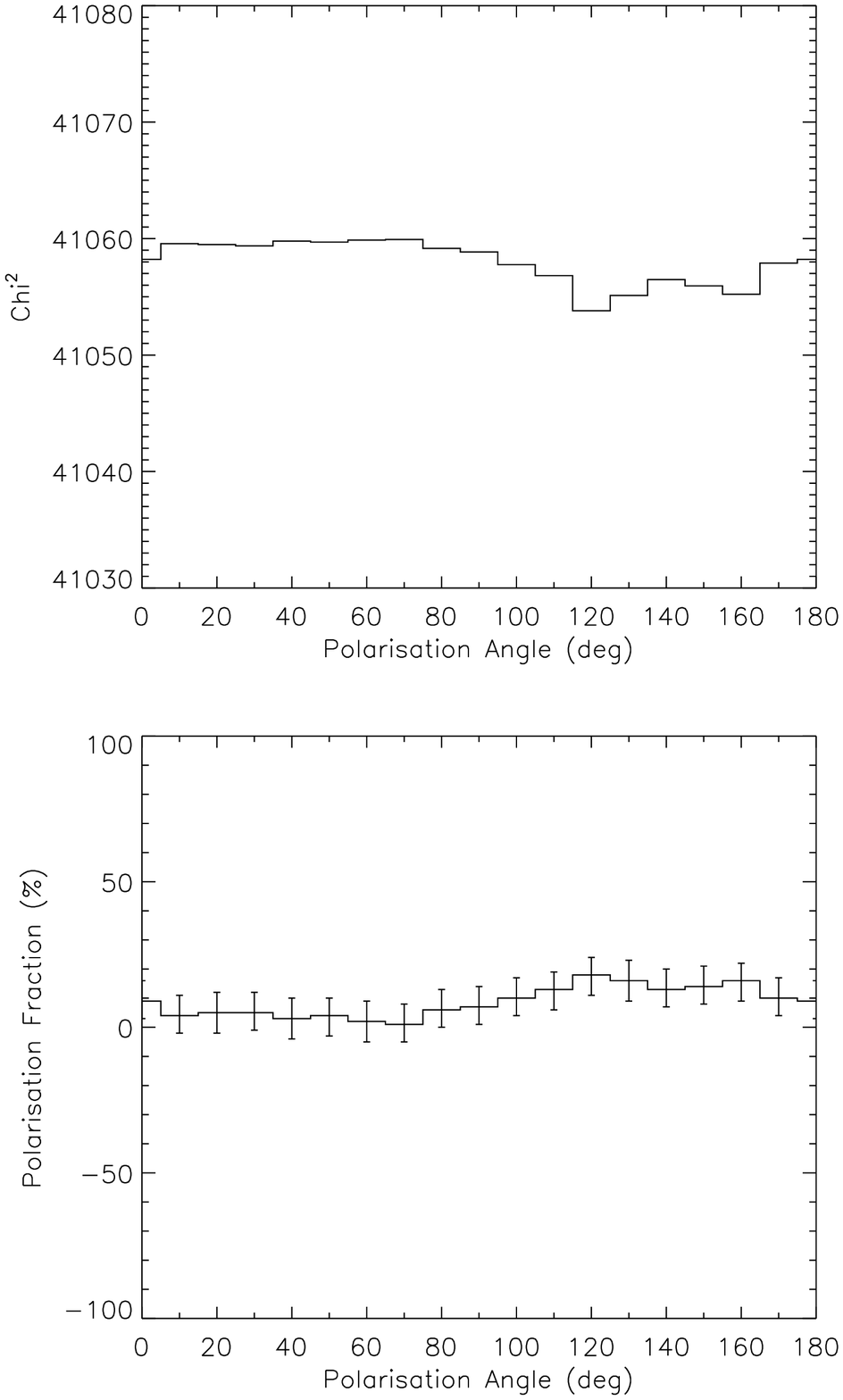}\\
\end{array}$
\end{center}
\caption{The same as Figure \ref{fig:mean}  for the 130-230 keV band.}
 \label{fig:LE}
\end{figure}

\begin{figure}[h]
\begin{center}$
\begin{array}{cc}
\includegraphics[width=2.4in]{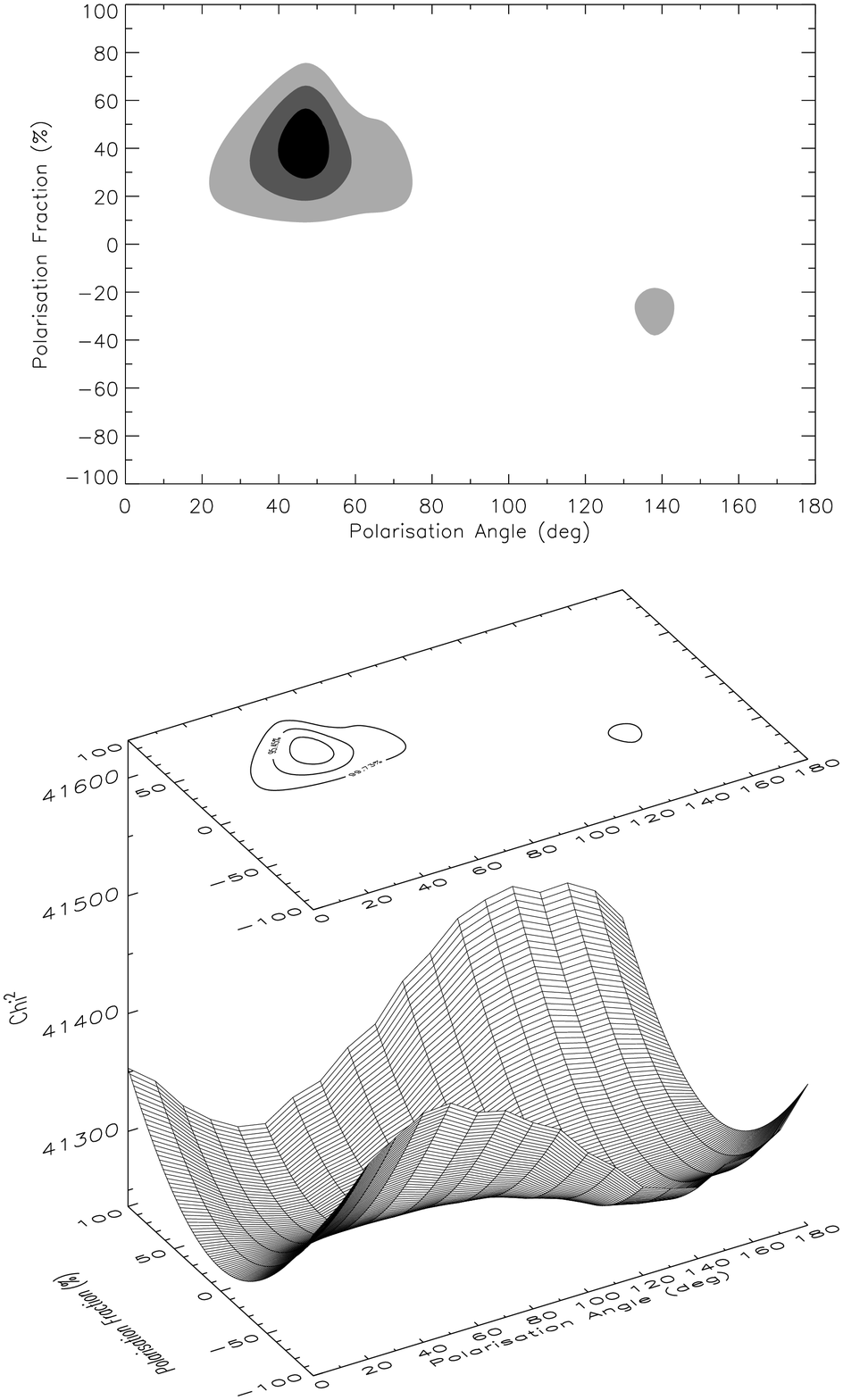}&
\put(-180,140){a)}
\put(-180,290){b)}
\put(-5,290){d)}
\put(-5,140){c)}
\includegraphics[width=2.5in]{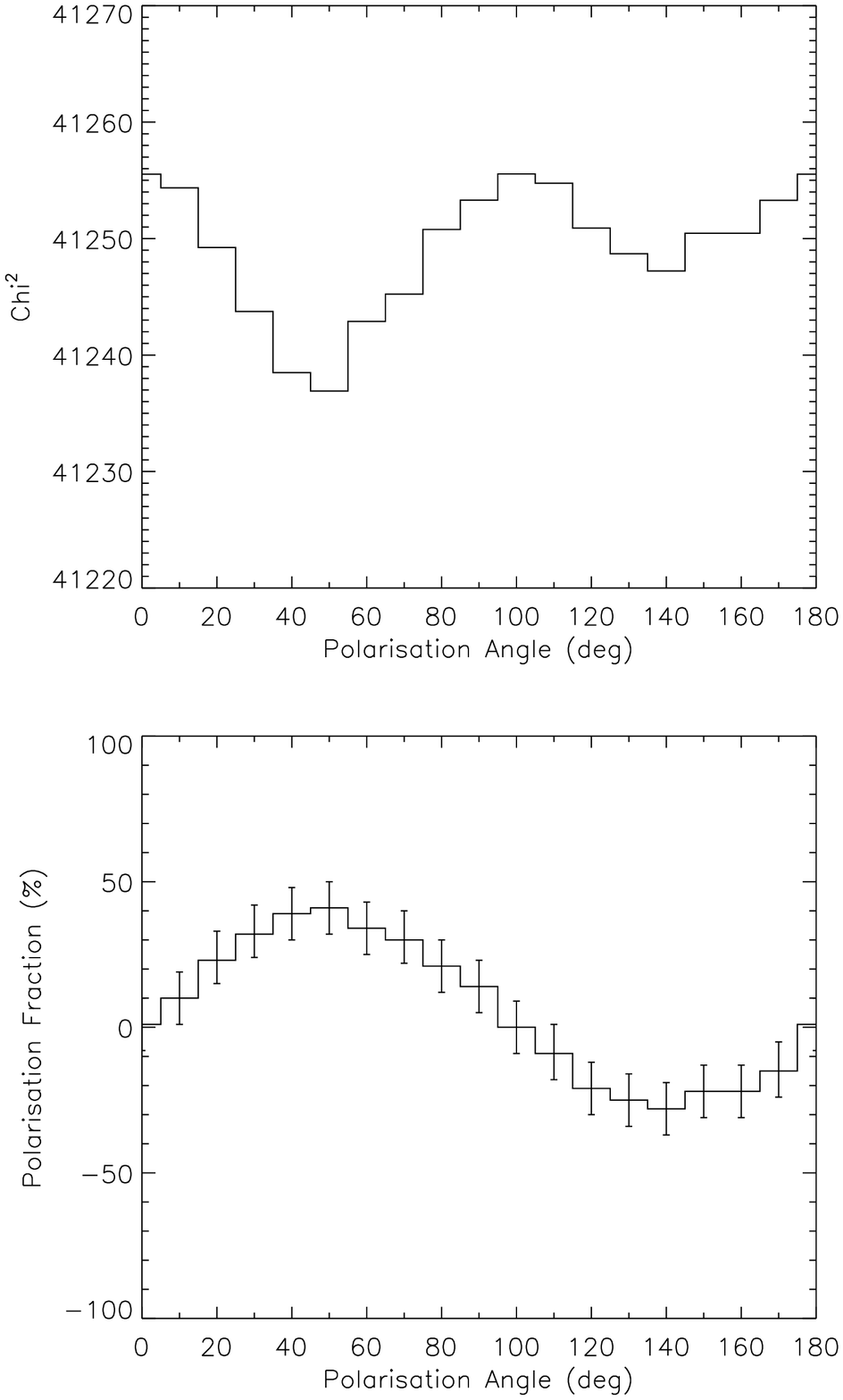}\\
\end{array}$
\end{center}
\caption{The same as Figure \ref{fig:mean}  for the 230-370 keV band.}
 \label{fig:ME}
\end{figure}

\begin{figure}[h]
\begin{center}$
\begin{array}{cc}
\includegraphics[width=2.4in]{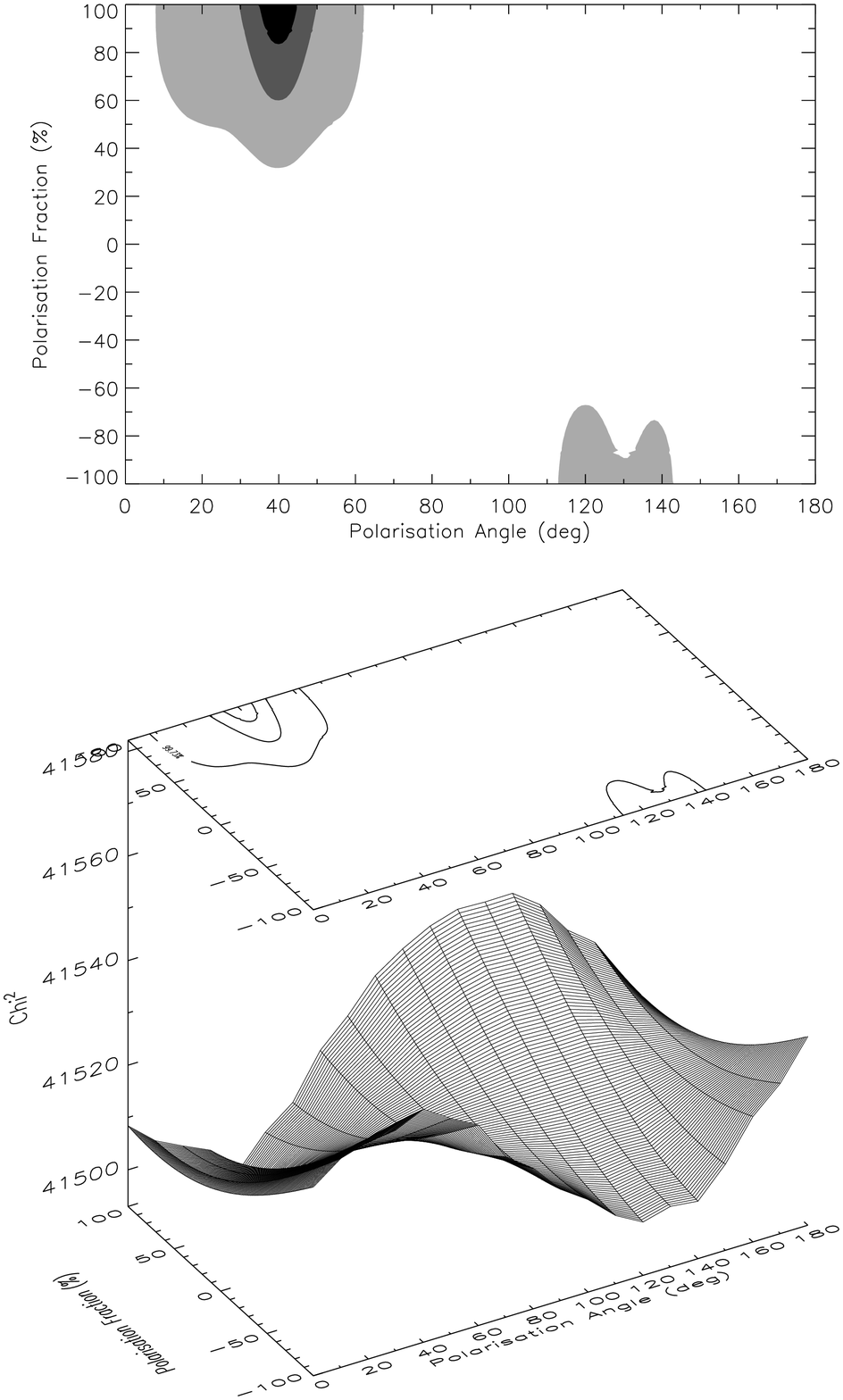}&
\put(-180,140){a)}
\put(-180,290){b)}
\put(-5,290){d)}
\put(-5,140){c)}
\includegraphics[width=2.5in]{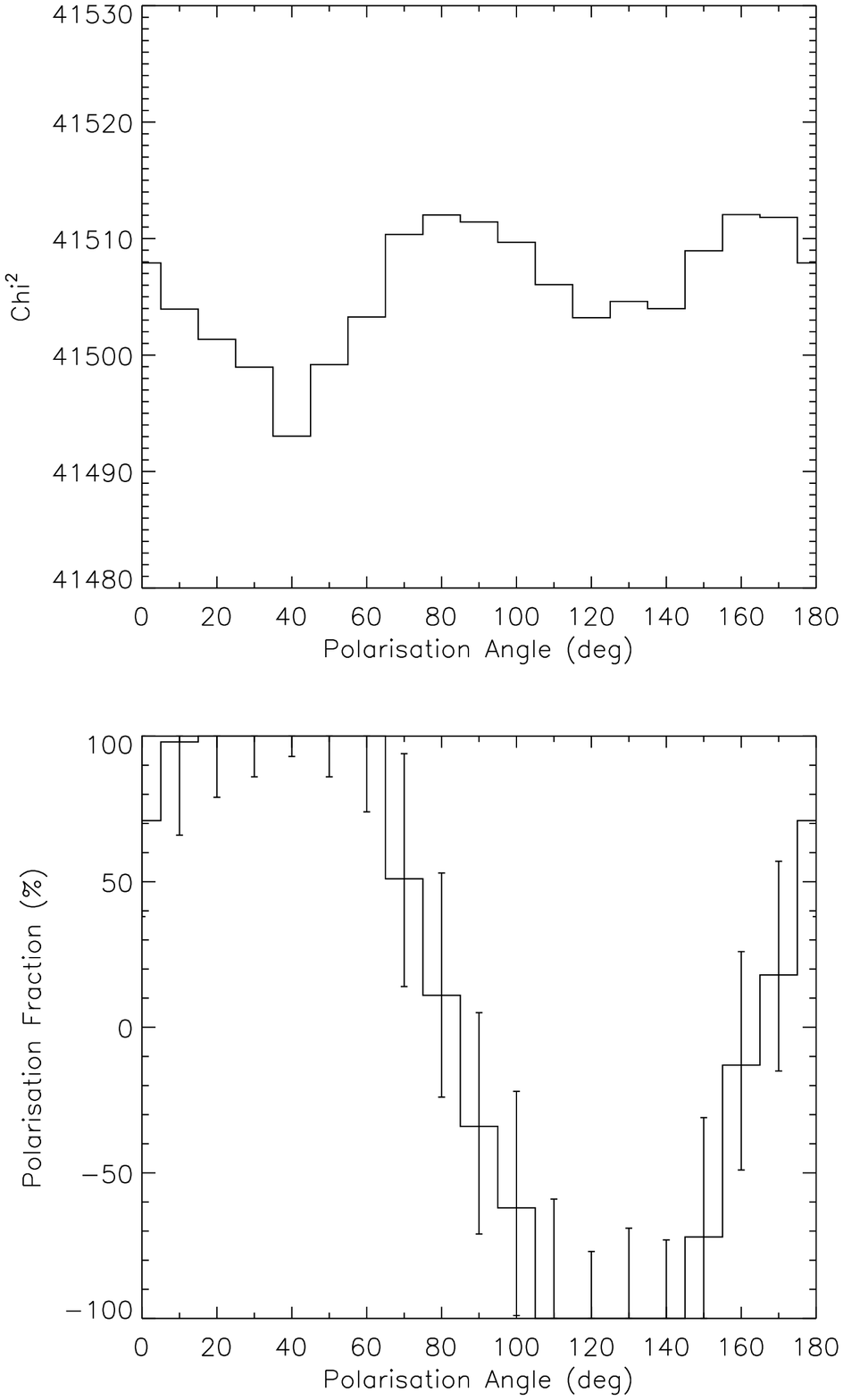}\\
\end{array}$
\end{center}
\caption{The same as Figure \ref{fig:mean}  for the 370-850 keV band.}
 \label{fig:HE}
\end{figure}

We have also performed the same analysis in three energy bands:
130 keV-230 keV, 230 keV-370 keV and 370 keV-850 keV.
The results are presented on Figures \ref{fig:LE},  \ref{fig:ME} and \ref{fig:HE}. \\
In the first energy range, the $\chi^2$ surface is quite flat, implying that no significant
 polarization is detected. 
The situation appears quite different when looking at higher energy.  
In the 230-370 keV and 370-850 keV  bands, the $\chi^2$ surfaces present  a clear
minimum around a polarization angle of $40^\circ - 50^\circ$, with an increasing polarization 
fraction.
The contour plots (top panels) give the best fit values together with their errors.
The position angles are rather well determined: 47$^\circ$ $\pm$ 4$^\circ$ and $39^\circ \pm 
3^\circ$,
in the medium and high energy band respectively. The polarization fraction is less easy to 
constrain. 
Nonetheless, we clearly see  that it  increases with energy, being undetectable at low energy,
(3 $\sigma$ upper limit of 20 \%, assuming the position angle determined above), 
it reaches $41 \% \pm 9\%$  in the medium band, and is constrained to be greater than
 75 \% at a 2 $\sigma$ level at higher energy.  

These results show that the high energy part of the Cyg X-1 emission is polarized, with 
 a polarization fraction increasing with energy. 
More precisely, we find that the  polarization is present down to $\sim$ 230 keV, whilst it
becomes hard to detect below that energy.
This implies that the hard X-ray/soft $\gamma$-ray emission
presents different properties   below and above $\sim$ 230 keV. This fits with the spectral 
behaviour seen in Cyg X-1, where a component in excess of the thermal Comptonization
appears above $\sim$ 250 keV. It is straightforward to link these two results and associate 
the
second component to the polarizing mechanism. Given the
high polarization fraction observed, a likely scenario relies on the synchrotron radiation
in a very ordered magnetic field as the mechanism responsible of  the high energy tail. 

\begin{figure} 
\plotone{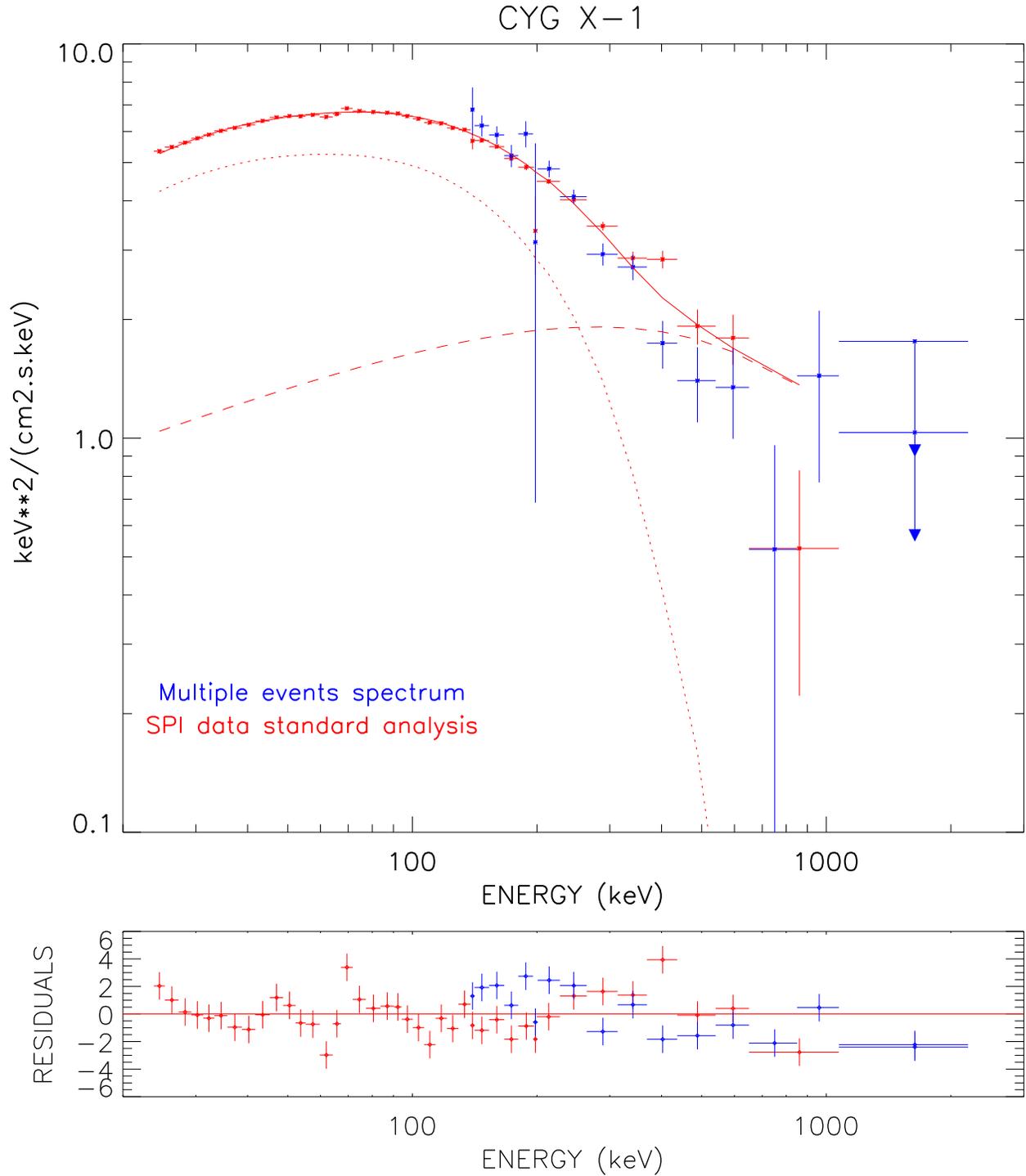}
\caption {The stacked spectra of SPI  obtained  from standard analysis (single events;
22 keV-2 MeV) and from multiple events (130 keV-8 MeV).
The solid lines represent the best fit model composed of a thermal 
Comptonization (reflec*CompTT, dotted curve) plus a fixed cutoff power-law model 
(photon index =1.6, $E_{cut}$ = 700 keV; dashed line). 0.5  \% of systematic  
have been added to the data.}
 \label{fig:spectreSM}
\end{figure}

This information can be introduced in the description of the observed spectral shape.
We combine the spectrum extracted from the single events (standard analysis), which provides 
the low energy emission (down to 22 keV), and the spectrum from the multiple events. A two 
component model is fit to the spectra using Xspec11. The first component, a thermal 
Comptonization model, Comptt (plus reflection) is used to describe the low energy part. The 
high energy component is represented by a cutoff power law, tentatively identified with the 
synchrotron process.
 Since the limited significance of the high energy data prevents us  actually constraining
the corresponding parameters, we have fixed  the cutoff power law with a photon index to 1.6,
 and the  cutoff value to $\sim$ 700 keV. These hypotheses result in a good 
 reproduction of the global spectrum.
 Fig \ref{fig:spectreSM} displays the resulting breakdown of the observed spectral emission,
in the proposed framework.
 Both components intersect around 240 keV, in agreement with a scheme where a polarized 
radiation
 emerges above this energy while the non polarized Comptonized flux dominates 
 at lower energy.\\

\section{DISCUSSION}
 
The recent detection  by INTEGRAL IBIS (Laurent et al. 2011) of a strongly polarized emission
above 400 keV in the X-ray binary Cyg X-1 emission can be seen as the beginning of a new era 
for 
the X-ray binaries high energy emission topic.  
Our results, based on SPI data, confirm the polarization of the high energy emission, 
and allow  a more precise picture. 

We find that the emission presents a strong polarization above 230 keV
with a position angle of $42^\circ$  $\pm$ $3^\circ$ East from the North. 
We have to note that this value is clearly
 incompatible with the IBIS/Compton mode  result reported by Laurent et al (2011) of 
$140^\circ \pm 10^\circ$.  However, due to an error in the angle convention, the correct value 
for the IBIS /Compton mode is
 $180 - 140^\circ= 40^\circ$, (P. Laurent, private communication), in perfect agreement with 
that presented here. \\
Both analyzes report an increase of the polarization fraction. 
In the SPI data, while only upper limits are accessible  below 230 keV ($f_{p} <$ 20\% at a 
3$\sigma$ level), the emission at higher energy presents a 
 strong polarization fraction : 40 $\pm$ 10 \% and then  $ >$  75 \% above 400 keV. 
High degrees of polarization  are usually expected from a synchrotron process. 
Generally, polarization requires a very ordered configuration with a single axis geometry or 
collimated structure.\\
From radio observations, Cyg X-1 is known to present some persistent jet emission, when in the 
hard state,
(Stirling et al. 2001), as well as transient emission (Fender et al. 2006).
The reported jet orientations are around -20$^\circ$ or -25$^\circ$, $\sim 65 ^\circ$ away 
(onto the sky),
from the polarization angle deduced for the high energy photons.
 These values are of prime importance to investigate  
the acceleration process and better comprehend the origin of the photons.

Combined with the spectral analysis, the polarization measurements provide crucial 
extra information in the global picture of the hard X-ray emission of Cyg X-1.  
 It has been known for a long time that
the spectral shape observed at high energy for Cyg X-1 requires the presence of two 
components.
Similarly, the evolution of the polarization fraction  with energy is expected in a scenario
 where  two mechanisms  coexist in the hard X-ray/soft $\gamma$-ray 
domain and intersect around 2-300 keV. We can establish a strong link between the second 
component and the polarizing mechanism, and suggest that in this source, the thermal 
Comptonization, unpolarized, component largely dominates
 the global emission up to 200 or 300 keV before  giving way to a second, polarized, 
 process.   
The statistics limit  the precision of our analysis but it is interesting to note that,
in principle, precise polarization measurements would allow both mechanisms to be 
disentangled,
if we assume that one is totally polarized (or have a high enough energy independent 
polarization fraction). 
Conversely, if the polarization fraction behavior is intrinsic to the emission mechanism,
its study will help constrain it.

The synchrotron process represents the most probable way to produce such strongly
 polarized emission. A scenario based on optically thin synchrotron emission
 from a partially self absorbed jet has already been  studied by
Zdziarski et al. (2012).
They fail to easily reconcile their model with the data, because they  impose a unique
broken power law to explain the source emission from radio to high energy. However, the
SPI data cover a 6.5 year period, while the radio or IR observation are very short. We 
therefore consider SPI data alone to estimate the parameters in the synchrotron framework. 
Even if it is difficult to put firm constrains, some rough features can be deduced.
In particular, the emitted photon energy distribution is directly connected to the electron
population parameters (see Ginzburg and Sirovatski, 1965). Similarly to Zdziarski et al.
 (2012), we find that a rather hard (photon) slope is needed, 
  since its contribution should decrease
 toward X-ray band. Finally, a good description of the second (synchrotron) component can 
 be obtained with a photon index of 1.6, corresponding to the reasonable value of 2.2 
 for an electron power law population. 
More information is potentially hidden in the observed cutoff energy (although not very 
constrained). 
We have fixed a cutoff value of 700 keV, slightly less than the folding energy
obtained by Zdziarski et al. (2012), which overestimated the MeV emission.
While the photon-photon absorption cannot be excluded, such a feature is more probable due to
the similar
 rollover in the emitting electron distribution. However, a more precise determination
 depends on the magnetic field strength, and requires some additional inputs or/and deeper
 modeling work.\\   
 A point worth mentioning is that  the detection of such a signal implies
its stability over long periods, and consequently, the same stability of the emitting medium. 
Whatever the mechanisms at work, they must keep (or come back to) the same characteristics.
This should include the process behind the relativistic 
electron acceleration.
  
\section{ SUMMARY AND CONCLUSIONS}

  We  have presented   our analysis into 
  the polarization phenomenon, through its role in the hard X-ray/soft $\gamma$-ray 
of X-rays binaries. 
The analysis of the multiple events in the observations of Cyg X-1 from the SPI instrument 
 refines (and corrects) the first result reported by Laurent et al. (2011) from the 
IBIS/Compton mode.
It shows that the source  emission is strongly polarized above $\sim$ 200 keV, with the 
fraction of 
polarization increasing with energy up to 850 keV.  
This behavior can be understood
in terms of two  emission mechanisms, one polarized and the other not, which contribute 
simultaneously
to the hard X-rays. The non polarized process would 
dominate  at low energy, cutting off above $\sim$150 keV and is well explained by thermal 
Comptonization. The polarized emission presents a harder spectral shape providing higher 
energy photons up to $\sim$ 1 MeV. 
This consolidates the  picture emerging from all high energy instruments that could not be 
reconciled with a single Comptonization law. 

Building on this idea, we have performed a spectral study from 20 keV to 1 MeV, gathering
the simultaneous standard SPI data (single events) and the multiple events analyzed in this 
paper.
At low energies, the Comptonization of the  disk photons by hot electrons explains both the 
spectral shape and the 
low/null polarization level of the emission.
The excess of emission observed at high energies above this thermal Comptonization law 
 can be associated with the polarization phenomenon and may be described by a cutoff power law 
 of photon index  1.6 with a cutoff energy (poorly constrained) around 700 keV.\\
Since the polarization requires a very organized medium to exist, a highly 
polarized emission is generally associated to a jet. Thus it is  likely that the  
high energy tail observed in the Cyg X-1 spectrum  is not associated to the disk but
to an electron population accelerated somewhere in the jet,  now well established to exist in  
this source
from radio observations. The position angle deduced from the analysis
 ($42^\circ$  $\pm$ $3^\circ$) is oriented $\sim$ $60^\circ$ away
  from the radio jet direction. 
An important point to note is that the polarization detection, particularly at
such a high degree, implies
that the phenomenon is persistent on a timescale of weeks to months.

To conclude, the detection of a polarized
  high energy emission in Cyg X-1 significantly impacts our view of  X-ray binaries.
A new scenario is now preferred where the jet structure that has been shown to exist in this 
class of source, can now be shown to perform a major role in the observed emission in the hard 
X-ray/soft $\gamma$-ray domain, a feature in many jet based models.\\
This new window opened into our understanding of X-ray binaries also strengthens the link 
between stellar mass black hole binaries and AGN characteristics. This will enable us to learn 
a lot from studies of both Galactic and extragalactic populations. Polarization measurements 
of many more sources will be required to build upon this picture. Measurements are strongly 
limited by the capabilities of the current generation of instruments which were not designed 
with polarimetry in mind. So to significantly progress this field, future missions should be 
planned to include a polarization ability that can help us understand the high energy 
environment in astronomical objects and break degeneracies between emission mechanisms in 
theoretical models. 
\\

\section*{Acknowledgments}  The \textit{INTEGRAL} SPI project has been completed under the
  responsibility and leadership of CNES.
   We are grateful to ASI, CEA, CNES, DLR, ESA, INTA, NASA and OSTC for support.
   MC and DC gratefully acknowledge financial support provided by the CNES, and we thank 
   the anonymous referee for interesting exchanges.


\begin{deluxetable}{lccccccc}
\tablewidth{0pt}
\tablecaption{Log of the \textit{INTEGRAL} SPI observations of Cyg X-1 used in this paper.}

\label{tab:revol}
\tabletypesize{\scriptsize}
\tablehead{
 \colhead{}\\
\colhead{revol }
&\colhead{Start} 
&\colhead{End} 
&\colhead{useful}\\
\colhead{number}
&&&\colhead{duration (ks)}\\

&\colhead{} \\
 
 }
\startdata

 79-80 (5x5)& 2003-06-07 00:59  &2003-06-12 03:35   & 293         \\
210-214 (A)& 2004-07-03 00:01 &2004-07-17 00:25  &  709        \\
251-252 (A)& 2004-11-03  14:23  & 2004-11-07 16:26   &  176   \\
259  \& 261 (H) & 2004-11-26 12:28 	 & 2004-12-03 15:43    &  143  \\
628-631 (A) & 2007-12-04 19:05  &  2007-12-15 21:08  &  388     \\
673 (A) &  	2008-04-18 17:41  &2008-04-19 22:09   &  54      \\
682-684 (A)  & 2008-05-14 08:13 & 2008-05-22 19:54    & 304        \\
803-806 (A) & 2009-05-11 08:27  &2009-05-22 11:32  &  371    \\
875(H*) \& 877(H)    & 2009-12-12 16:18  & 2009-12-19 20:57    & 160     \\

\enddata

\tablecomments{In the first column, the letter after the revolution number indicates the 
dithering strategy used:
(5x5) for the standard 5X5 pattern (see section 2); (A) for a pointing strategy centered 
between Cyg X-1 and Cyg A region;
(H) for the hexagonal pattern and (GP) for a Galactic Plane scan. (H*) During the rev 875, the 
pointing strategy 
follows a pattern proposed by Wilms et al in their AO-7 proposal. All this information
is available on the dedicated ESA site web http://integral.esa.int/isocweb.
}
 
\end{deluxetable}

\end{document}